\documentclass[epj]{svjour}
\usepackage{amssymb}
\usepackage{amsmath}
\usepackage{array} % to improve presentation of tables
\usepackage{epsfig}
\usepackage{graphics}
\usepackage{colortbl}
\usepackage{hhline}
\usepackage{multirow}
\begin{document}
\title{Hadron multiplicity induced by top quark decays at the LHC.}
\author{R.A.~Ryutin\thanks{\emph{e-mail:} Roman.Rioutine@cern.ch}\inst{1}
}                     % Do not remove
\institute{{\small Institute for High Energy Physics},{\small{\it 142 281}, Protvino, Russia}}
\abstract{
The average charged hadron multiplicities induced by top quark decays are
calculated in pQCD at LHC energies. Different modes of top 
production are considered. Proposed measurements can be used as an
additional test of pQCD calculations independent on a fragmentation model.
 % end of PACS codes
} %end of abstract
\authorrunning
\titlerunning
\maketitle

\section{Introduction}
\label{intro}
The study of unstable heavy particles like W, Z bosons, top quarks 
and others (arising in different extensions of the Standard Model)                                                                                                                                                is one of the leading directions in the modern high energy 
physics. To determine particle parameters (charge, mass, width, decay 
modes etc.) we have to look deep inside their production and 
decay mechanisms.

In this article processes with top production are considered due
to its specific properties. Top quark is extremely elusive 
object. Because its mass is so large 
($m_t^{exp}=172\pm 0.9\pm 1.3\;{\rm GeV}$)~\cite{PDG2010}, it
can decay into on-shell W-bosons, i.e., the two-particle decay mode $t\to bW^+$
is kinematically possible. The SM predicts the top quark to decay almost exclusively
into this mode. The on-shell W-boson can then decay leptonically
or hadronically with coupling strengths given by the
Cabibo-Kobayashi-Maskawa (CKM) matrix. The top quark
decay proceeds extremely fast, in less than 
$\tau_t=1/\Gamma_t\simeq 5\times 10^{-25}\;{\rm s}$, which
is shorter than the time scale to form hadrons 
$\tau_{had}\simeq 1/\Lambda_{QCD}\approx 3\times 10^{-24}\;{\rm s}$ and
almost 13 orders of magnitude smaller than the lifetime of hadrons, which involve 
the next heaviest quark, $\tau_b\simeq 1.5\times 10^{-12}\;{\rm s}$. The width
$\Gamma_t$ acts as a physical "smearing", and the top production becomes
a quantitative prediction of pQCD, largely independent on nonperturbative
phenomenological algorithms. That is why top measurements are directly related 
to pQCD tests.

The principal observables used in tests of pQCD are typically 
measurements of jets, high transverse momentum particles and 
event shapes. There are relative advantages and disadvantages
in using these observables. Jet measurements are expected to 
have a close correlation in direction and momentum with the 
parton which gave rise to it. However, several complications
concerning jet definitions (algorithms) arise when using jet 
measurements.  While the infrared and collinear safety of event 
shapes allows safe perturbative predictions, we have to take 
into account resummation of large logarithms.  Evolution
of structure functions can be predicted by pQCD calculations, but we 
have to consider different kinematical 
regimes (see, for example, recent review on QCD tests~\cite{QCDtests}). The 
transition from partons to hadrons cannot be accomodated within
perturbative QCD. Fragmentation mechanism is usually simulated
by the use of additional approaches (string fragmentation, cluster
fragmentation). It was shown~\cite{PK1,PK2} that measurements of average
charged hadron multiplicitiy in a jet (especially jet produced in 
heavy quark decay) can serve as a precise test of 
pQCD \emph{\bf independent on a fragmentation model}.

Measurements of quark and gluon jets 
show visible differences between them. For example, gluon jets are 
fatter, softer and have higher multiplicity~\cite{ellis1988}. In events
induced by heavy quark jets the hadron multiplicity is smaller than in analogous
events triggered by light quark jets~\cite{picsmult1}-\cite{picsmult3}. The
situation is similar to the classical theory, where the more is the mass of a charged
particle, the less intensive is the radiation from it. The dependence of the multiplicity 
in a quark jet on the "primary quark" mass was observed in 
previous experiments, in particular, at the SLC, KEK, LEPI and LEPII (see~\cite{PK3} and references therein). Leading 
order pQCD predicts "specific scaling"~\cite{PK1},\cite{PK3}-\cite{scaling1}, i.e. energy independent difference 
between average charged multiplicities of light and heavy quark jets.
The advantage of multiplicity measurement is that we do not need large number of events,
and it can be used in rare processes like, for example, single top production.

The QCD radiation associated with $t\bar{t}$ production has been 
treated earlier in~\cite{ttrad1,ttrad2}. These papers consider the effect,
that gluons are also radiated from the b's from t-decay. This effect has
been also taken into account here. Recently top production at the 
ILC $e^+e^-$ collider was considered~\cite{PK2}. 
Following the basic idea of Ref.~\cite{PK2} it is possible to analyse the
case of top production at the LHC. Since in pp events the energy of
parton interaction is not fixed
and changes in the range from about 400~GeV up to several TeV, we can study the 
evolution of average multiplicities in this energy domain. Of course, several obvious 
difficulties can arise (color interconnection, interference effects). It is shown, that 
there are some cases where we can successfully avoid the above complications.

The paper is organized as follows. In the first section 
there is an outlook of the basic model. The second one is devoted
to numerical results for different mechanisms of the top production at the 
LHC. Complicated formulae and calculations are collected in Appendices.

\section{Model}
\label{sec:1}
%fig:pr
\begin{figure}[tb!]
\begin{center} 
\resizebox{0.5\textwidth}{6cm}{%
  \includegraphics{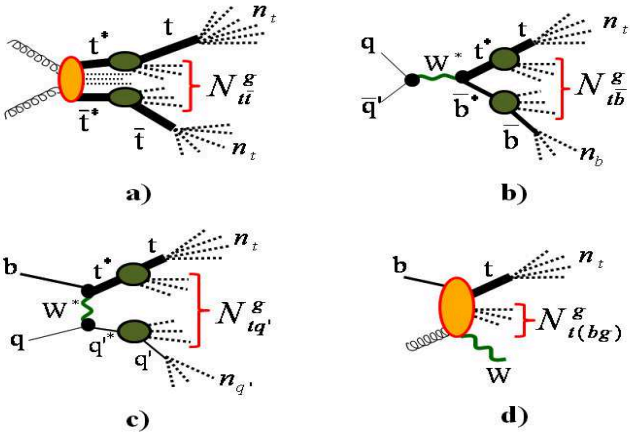}
}
\caption{\label{fig:pr} Parton level processes of top production at LHC considered in 
this paper: a) dominant $t\bar{t}$ production in the gluon-gluon fusion; b) s-channel 
single top production; c) t-channel single top production; d) tW production.}
\end{center} 
\end{figure}
The model used in this section is analogous to the one presented 
in~\cite{PK1}. In this article we consider processes of
the type $p+p\to t\bar{t}+X$ or $p+p\to t+X$ (see Fig.~\ref{fig:pr}). The basic 
formula for inclusive production of the system $M$ in the collinear 
approximation looks as follows
\begin{equation}
\label{cspptoMX}
\frac{d\sigma_{pp\to M\;X}(s)}{dx_1 dx_2 d\Phi_M}=
\sum_{i,j}f_i(x_1)f_j(x_2)\frac{d\hat{\sigma}_{ij\to M}}{d\Phi_M}(x_1x_2s;\ \{\Phi_M\}),
\end{equation}
where $f_i(x)$ is the probability to find parton $i$ (quark, anti-quark or gluon) 
with the longitudinal momentum $x\sqrt{s}/2$ in a proton. Renormalization 
and factorization scales are hidden in 
$f$ and $\hat{\sigma}$. Here M is the system of different final states like  
$t\bar{t}$, $t\bar{b}$, $tq$, $tW$, which 
corresponds to different mechanisms
of the inclusive top production at the LHC. $X$ 
includes beam remnants ($X_{beam}$) and also the 
secondary radiation induced by color interactions 
inside M ($X_M$) plus possible interaction between M and beam remnants ($X_{M-beam}$), if 
M is not a color singlet. Variables $x_{1,2}$ are fixed in every 
separate event and can be calculated experimentally. Usually the sum in~(\ref{cspptoMX}) can 
be approximated by a single factorized term 
which includes parton distributions multiplied by
the amplitude squared of parton-parton cross-sections:
\begin{eqnarray}
\label{prtt} && g+g\to t\bar{t}+X_M \mbox{(dominates  
at the LHC)},\\
\label{prt1} && q+\bar{q}'\to W^*\to t\bar{b}+X_M,\\
\label{prt2} && b+q\to tq'+X_M,\\
\label{prt3} && b+g\to tW+X_M,\\ 
\label{prt4} && q+g\to t\bar{b}q'+X_M,\\
\label{prt5} && g+g\to t\bar{b}W+X_M,
\end{eqnarray}
where $q,\; q'$ denotes corresponding light quarks. For our purposes 
it is enough to consider initial parton collisions instead
of pp process, since we have to calculate
only the charged hadron multiplicity of the $M$ plus $X_M$ in a separate 
event. Complications concerning $X_{beam}$ and $X_{M-beam}$ are discussed below.
$X_M$ comes from virtual gluon radiation.

For the average multiplicity of hadrons in $M+X_M$ we can use the expression
similar to Eq.~(5) in~\cite{PK1}:
\begin{eqnarray}
&& N_{M+X_M}^h(Q^2) = n_M + \nonumber\\
&& \!\int \! \frac{d^4k}{(2\pi)^4} \,
\Pi_{\mu \nu}^{a b}(q_1,q_2,k) \, d^{\, \mu \alpha}_{a a'}(k) \, d^{\,
\nu \beta}_{b b'}(k) \, n^{a' b'}_{\alpha \beta}(k) \;,\label{hadron_mult}
\end{eqnarray}
where $d^{\, \mu \nu}_{a b}(k)\equiv i\;\tilde{d}^{\, \mu \nu}_{a b}(k)/(k^2+i0)$ is the propagator of the gluon
with momentum $k$, $q_{1,2}$ are momenta of initial partons. Here 
and below $(a, b)$ and $(a', b')$ denote
color indices, $Q=\sqrt{(q_1+q_2)^2}$ is the energy of colliding partons. 

The first term in the r.h.s. of Eq.~(\ref{hadron_mult}), $n_M$, is the
multiplicity from the fragmentation of leading particles in the final 
state. For example, in the process~(\ref{prtt}) $n_M=n_{t\bar{t}}=2n_t$, where
$n_t$ was calculated in~\cite{PK2}. In other processes $n_M$ is appropriate 
combination of multiplicities which are taken from the analysis of data 
and pQCD calculations:
\begin{eqnarray}
&& \label{MULTnth} n_t^H\equiv n_t(t\to{\rm hadrons})=41.03\pm 0.54\mbox{\cite{PK2}},\\
&& \label{MULTntl} n_t^L\equiv n_t(t\to l\bar{\nu}_l+{\rm hadrons})=21.9\pm 0.53\mbox{\cite{PK2}},\\
&& \label{MULTnW} n_W(W\to{\rm hadrons})=19.34\pm 0.10\mbox{\cite{PK2}}\\
&& \label{MULTnbc} n_c=2.6,\; n_b=5.5\mbox{\cite{dataMULTbc}},\; n_q=1.2\mbox{\cite{dataMULTq}}.
\end{eqnarray}

Tensor $n^{a' b'}_{\alpha \beta}(k)$ is given in~\cite{PK1}:
\begin{equation}\label{gluon_frag}
n^{a' b'}_{\alpha \beta}(k) = \left( -g_{\alpha \beta} \, k^2 +
k_{\alpha} k_{\beta} \right) \delta^{a' b'} n_g(k^2) \;,
\end{equation}
where dimensionless quantity $n_g(k^2)$ describes the average
multiplicity of hadrons in the gluon jet with the virtuality
$k^2$. It is, of course, gauge invariant, and depends only on the
virtuality $k^2$. 

The quantity $n_g(k^2)$ cannot be calculated perturbatively. It is
usually assumed that the average hadron multiplicity is
proportional to $n_g(k^2, Q_0^2)$, i.e. the average multiplicity
of (off-shell) partons with the ``mass'' $Q_0$ (the so-called
local parton-hadron duality):
\begin{equation}\label{LPHD}
n_g(k^2) = n_g(k^2, Q_0^2) \, K(Q_0^2) \;,
\end{equation}
where $K(Q_0^2)$ is a phenomenological energy-independent factor.
The QCD evolution equations for both $n_g(k^2, Q_0^2)$ and
\begin{equation}
\label{Ngdefine}N_g(k^2, Q_0^2)=\int\limits_{Q_0^2}^{k^2}\frac{dp^2}{p^2}n_g(p^2,Q_0^2)
\end{equation} 
are derived in~\cite{PK1}, and also in old 
works~\cite{eqNg1,eqNg2}. We use the 
``conventional standard'' value $Q_0=1\;{\rm GeV}$ for further 
numerical calculations. Here $N_g(k^2,Q_0^2)$ is the average 
multiplicity from the gluon
jet whose \emph{virtuality $p^2$ varies up to $k^2$}. Very often 
$N_g$ is erroneously called the average
multiplicity of the gluon jet with \emph{fixed virtuality} $k^2$.
This meaning should be addressed to $n_g$ only. 

In this paper we will use two phenomenological 
expressions (to estimate theoretical uncertainties) 
for $n_g$ which can be found in~\cite{PK1}:
\begin{eqnarray}
\label{ng} n^i_g(k^2)&\!=&\!k^2\frac{d}{dk^2}N^i_g(k^2,Q_0^2),\\
  \!\!\!N^1_g(k^2)&\!=&\!3.89+0.01\exp\left[ 1.63\sqrt{\ln\left( \frac{k^2}{\Lambda_1^2}\right)}\right],\nonumber\\
\label{Ng1}\Lambda_1&\!=&\!0.87\;{\rm GeV}\mbox{\rm (QCD motivated)},\\
 \label{Ng2} \!\!\!N^2_g(k^2)&\!=&\!4.21+0.012\ln^2\frac{k^2}{\Lambda_2^2},\;\Lambda_2=0.93\;{\rm GeV}.
\end{eqnarray}
Parameters for these functions were obtained 
in~\cite{PK2} (see eq.(3) and Fig.8 in this reference, where $n_g$ 
corresponds to the function $N_g$ in the present paper) by
the fitting of the data from~\cite{scaling1}.

 In Eq.~(\ref{hadron_mult}) the first factor of the integrand is given by 
 \begin{eqnarray}
 &&\Pi_{\mu \nu}^{ab}(q_1,q_2,k) = \left[\prod_{i=1}^{N_{in.gl.}} \tilde{d}^{\rho_i \sigma_i}(q_i,n)\delta_{a_ib_i}\right] \times\nonumber\\
 &&\phantom{\Pi_{\mu \nu}^{ab}(q_1,q_2,k) = \left[\right.}\label{convolution}\Pi_{\{\rho_i \sigma_i\};\;\mu
 \nu}^{\{a_ib_i\};\;ab}(q_1,q_2,k) \;,
 \end{eqnarray}
where 
$\Pi_{\{\rho_i \sigma_i\};\;\mu \nu}^{\{a_ib_i\};\; ab}(q_1,q_2,k)$ can be calculated in the
first order in the strong coupling constant as the 
amplitude squared of the corresponding process~(\ref{prtt})-(\ref{prt5}) with $X_M=g$ normalized
to the total rate of the process without $X_M$. Sum in the product of polarization vectors of
initial gluons ($N_{in.gl.}=0,1,2$) forms usual factor (here we use the axial gauge since it 
simplifies much theoretical calculations)
\begin{eqnarray}
 \tilde{d}^{\rho_i \sigma_i}(q_i,n)&\!=&\!\sum_{\lambda=1,2} \epsilon_{(i);\lambda}^{\rho_i}\epsilon_{(i);\lambda}^{*; \sigma_i}=\nonumber\\
\label{axiald} -g^{\rho_i \sigma_i}&\!+&\!\frac{q_i^{\rho_i}n^{\sigma_i}+n^{\rho_i}q_i^{\sigma_i}}{qn}-\frac{n^2}{qn^2}q_i^{\rho_i}q_i^{\sigma_i},
\end{eqnarray}
where $\epsilon_{(i);\lambda}^{\rho}q_{i; \rho}=0$, $q_i^2=0$, $n$ is an appropriate four-vector
in the corresponding process (see Appendices B,C).

The quantity~(\ref{convolution}) satisfies the equality
\begin{equation}
\label{gaugeinvar}
k^{\mu} \Pi_{\mu \nu}^{ab}(q_1,q_2,k) = 0
\end{equation}
due to the \emph{\bf general theorem}~\cite{Predazzigauge}: {\sl If the QCD amplitude
is written}
\begin{eqnarray}
&&\!\!\!\!\!\!\!\!\!\!\mathcal{A}_{QCD} =\nonumber\\
&&\!\!\!\!\!\!\!\!\!\!\label{GIQCD}\epsilon_{\nu_1}^{\star}(\kappa_1) ...\epsilon_{\nu_N}^{\star}(\kappa_N)\, \mathcal{T}_{a_1 ...a_N \, ; \,\, b_1 ...b_M}^{\nu_1 ... \nu_N \,; \,\, \mu_1 ...\mu_M} \, \epsilon_{\mu_1}(k_1) ...  \epsilon_{\mu_M}(k_M)
 \end{eqnarray}
 {\sl then one gets zero if} \emph{any number},  $\geq 1$, {\sl of the polarization vectors} $\epsilon_{\mu_j}(k_j)$ {\sl and/or} $\epsilon^*_{\nu_i}(\kappa_i)$ {\sl are replaced by} $k_{j\, , \, \mu_j}$  {\sl and/or}  $\kappa_{i\, , \, \nu_i} $  {\sl respectively,} \emph{provided} {\sl that all these} $k_j${\sl 's and} $\kappa_i${\sl 's, with the exception of} \emph{at most one of them} {\sl, satisfy} $k_j^2 =0$ {\sl and} $\kappa_i^2=0${\sl.}

If we take into account Eq.~(\ref{gaugeinvar}) and introduce the function
\begin{equation}
\label{PIfun}
\Pi(Q^2,k^2,kq_1,kq_2)=(-g^{\mu \nu}) \delta_{ab} \Pi_{\mu \nu}^{ab}(Q^2, k^2,
kq_1,kq_2),
\end{equation}
the final formula for the multiplicity will look as follows:
\begin{eqnarray}
&&N_{M+X_M}^h(Q^2) =n_M+N^g_M\equiv\nonumber\\
&&\label{hadron_mult2}n_M + \!\int \! \frac{d^4k}{(2\pi)^4} \,
\Pi(Q^2,k^2,kq_1,kq_2) \left.\frac{d}{dp^2}N_g(p^2)\right|_{p^2=k^2}.
\end{eqnarray}
Concrete form of the function $\Pi$ for different processes can be found in
Appendices B,C.

\section{Numerical results of calculations}
\label{sec:2}

 In this section we consider numerical results for average charged
multiplicities in different processes of top production at the LHC. Below we consider the phase space when final jets have low transverse
momentum cuts $P_t$, and the final 
gluon jet can not be experimentally separated from one 
of final quark jets (i.e. gluon jet lies within 
the cone $\cos\theta_{gq}>R=0.9$, where $\theta_{gq}<0.45$ is the angle 
between the gluon and quark jets).
% tab:Ngtt 
\begin{table}[b!]
\caption{\label{tab:Ngtt} Multiplicity $N^g_{t\bar{t}}$ for different cuts of jet transverse momenta $P_t$ and
the energy of gluon-gluon collision.}
\centering
\begin{tabular}{|c|c|c|c|}
\hline
 {\cellcolor[gray]{0.9}$N^g_{t\bar{t}}(Q,P_t)$} & \multicolumn{3}{c|}{$Q$, GeV}  \\
 \hhline{|-|-|-|-|}
 $P_t$, GeV & 600  & 1000  & 1500 \\
 \hline
 10 & 2.82$\pm$0.07  & 8.29$\pm$0.2 & 15.75$\pm$0.36 \\
 \hline
 30 & 0.76$\pm$0.02 & 2.96$\pm$0.06 & 6.58$\pm$0.12 \\
 \hline
 50 & 0.4$\pm$0.01 & 1.3$\pm$0.02  & 3.63$\pm$0.05 \\
 \hline
\end{tabular}
\end{table}

\begin{figure}[t!]
\begin{center} 
\resizebox{0.5\textwidth}{5cm}{%
  \includegraphics{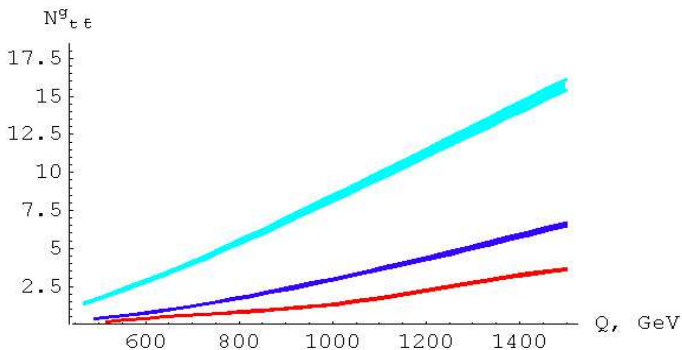}
}
\caption{\label{fig:Ngtt} $t\bar{t}$ production. Multiplicity $N^g_{t\bar{t}}$ versus $Q$ for different cuts of jet transverse momenta. Top-down: $P_t=10\;GeV\to P_t=30\;GeV\to P_t=50\;GeV$.}
\end{center} 
\end{figure}

Let us
begin with the inclusive $t$ $\bar{t}$ production~(\ref{prtt}). The total
cross-section of the inclusive
process $pp\to t\bar{t}+X$ is about 833~pb at 14~TeV. In this
article we consider only the gluon-gluon fusion
mechanism of this process since at LHC it is 
dominant. Numerical values 
for $N^g_{t\bar{t}}$ are given in the table~\ref{tab:Ngtt} and on 
the Fig.~\ref{fig:Ngtt}. Here and below theoretical errors
are estimated by the use of two different 
parametrizations~(\ref{Ng1}),(\ref{Ng2})
for the hadronic multiplicity in a gluon jet. The average charged multiplicity in different decay 
modes (hadronic, semileptonic and leptonic) can be calculated as follows
\begin{eqnarray}
 \label{mult:ttHH} N^h_{t\bar{t}\to hadrons}(Q)&\!=&\!2n^H_t+N^g_{t\bar{t}}(Q),\\
 \label{mult:ttHL} N^h_{t\bar{t}\to l\bar{\nu}_l+hadrons}(Q)&\!=&\!n^H_t+n_t^L+N^g_{t\bar{t}}(Q),\\
 \label{mult:ttLL} N^h_{t\bar{t}\to l^+l^-\nu_l\bar{\nu}_l+hadrons}(Q)&\!=&\!2n_t^L+N^g_{t\bar{t}}(Q).
\end{eqnarray}
As you see on the Fig.~\ref{fig:Ngtt}, the dependence of $N^g_{t\bar{t}}$ on the energy is visible. In this work 
we  assume that color reconnection of $t\bar{t}$ and beam remnants
is small due to the strong suppression of this processes with
high transverse momentum transfer (typical jet transverse momentum cut at the LHC
is about $20$-$40$~GeV). Also $t$ and $\bar{t}$ fragment
independently after the interaction inside the $t\bar{t}$ system. It
looks similar to the process of $W^+W^-$ fragmentation in $e^+e^-$ 
annihilation. The effect of possible color reconnection
was investigated by comparing hadronic 
multiplicities in $e^+e^-\to W^+W^-\to q\bar{q}'q\bar{q}'$
and $e^+e^-\to W^+W^-\to q\bar{q}'l\bar{\nu}_l$. No evidence
for final state interactions was found by
measuring the difference
$<n_{4q}^h>-2<n_{2ql\bar{\nu}}^h>$~\cite{WWfragm1},\cite{WWfragm2}.
The values for average charged multiplicities can be compared with the present LHC data
on the inclusive $t\bar{t}$ production.

% fig:Ngtt
% tab:Ngt1
\begin{table}[b!]
\caption{\label{tab:Ngt1} S-channel single top production. Multiplicity $N^g_{t\bar{b}}$ for different cuts of jet transverse momenta $P_t$ and
the energy of parton-parton collision.}
\centering
\begin{tabular}{|c|c|c|c|}
\hline
 {\cellcolor[gray]{0.9}$N^g_{t\bar{b}}(Q,P_t)$} & \multicolumn{3}{c|}{$Q$, GeV}  \\
 \hhline{|-|-|-|-|}
 $P_t$, GeV & 600 & 1000 & 1500 \\
 \hline
 10 & 11$\pm$0.32 & 14.8$\pm$0.41 & 18.7$\pm$0.5 \\
 \hline
 30 & 6.55$\pm$0.19 & 10$\pm$0.27 & 13.2$\pm$0.33 \\
 \hline
 50 & 4.33$\pm$0.12 & 7.55$\pm$0.2 & 10.4$\pm$0.25 \\
 \hline
\end{tabular}
\end{table}

\begin{figure}[t!]
\begin{center} 
\resizebox{0.5\textwidth}{5cm}{%
  \includegraphics{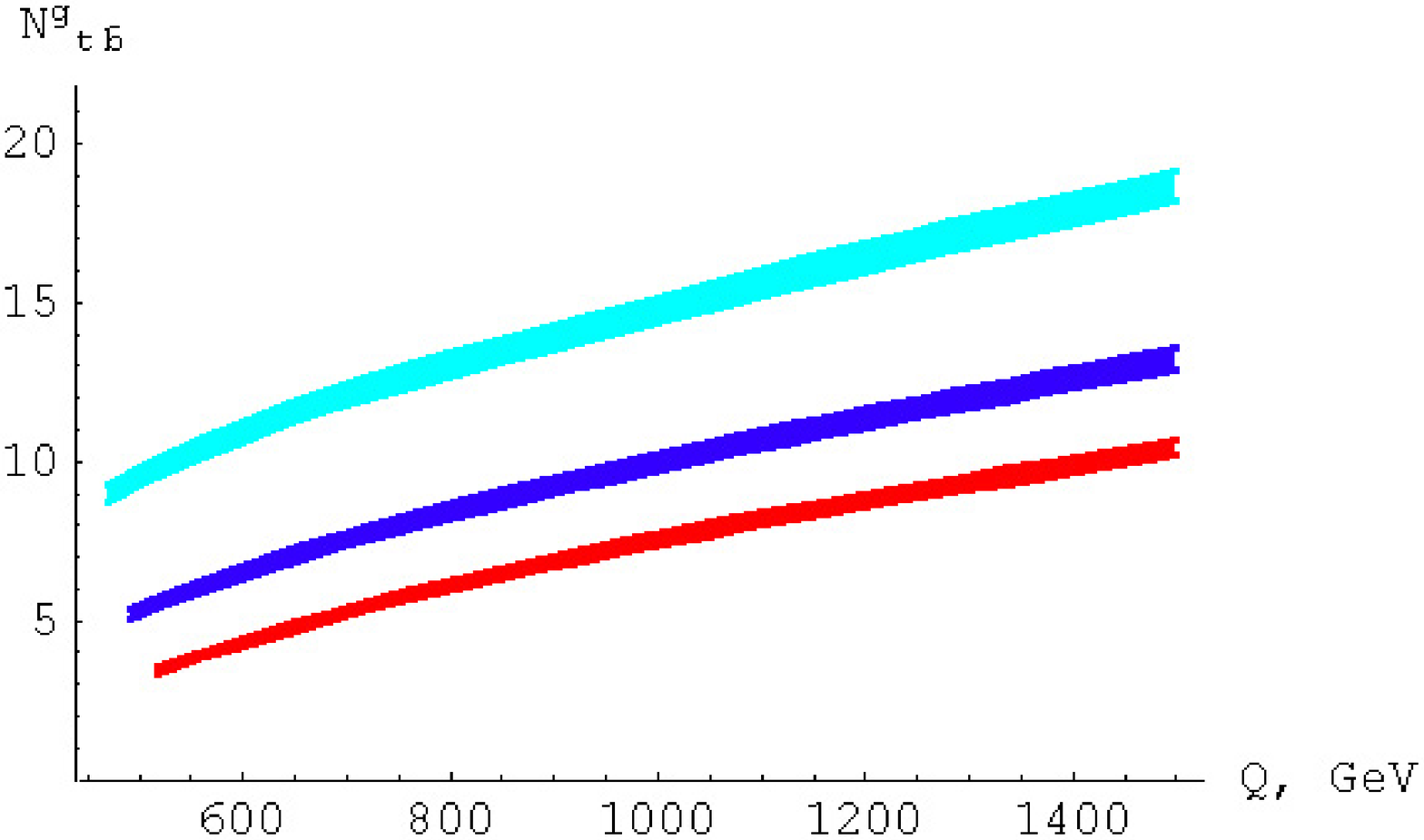}
}
\caption{\label{fig:Ngt1} S-channel single top production. Multiplicity $N^g_{t\bar{b}}(Q)$ versus $Q$ for different 
cuts of jet transverse momenta. Top-down: $P_t=10\;GeV\to P_t=30\;GeV\to P_t=50\;GeV$. }
\end{center} 
\end{figure}

 The case of s-channel single top production~(\ref{prt1})
is close to the $e^+e^-$ one, since the final state is
a result of $W$ decay, i.e. color singlet. That is why
we have no color reconnection with beam remnants. However, the 
cross-section of this process is rather small (about 11~pb at 14~TeV), and the
experimental task on the extraction of the multiplicity looks 
more difficult than, for example, in t-channel single top or $t\bar{t}$ 
production. Numerical values 
for $N^g_{t\bar{b}}$ are given in the table~\ref{tab:Ngt1} and on 
the Fig.~\ref{fig:Ngt1}. The average charged multiplicity in different decay 
modes can be calculated as follows
\begin{eqnarray}
 \label{mult:t1HH} N^h_{t\bar{b}\to hadrons}(Q)&\!=&\!n^H_t+n_b+N^g_{t\bar{b}}(Q),\\
 \label{mult:t1HL} N^h_{t\bar{b}\to l\bar{\nu}_l+hadrons}(Q)&\!=&\!n^L_t+n_b+N^g_{t\bar{b}}(Q).
\end{eqnarray}
The energy dependence is not so strong as in the previous case (see Fig.~\ref{fig:Ngt1}).

% fig:Ngt1
\begin{figure}[b!]
\begin{center} 
\resizebox{0.5\textwidth}{5cm}{%
  \includegraphics{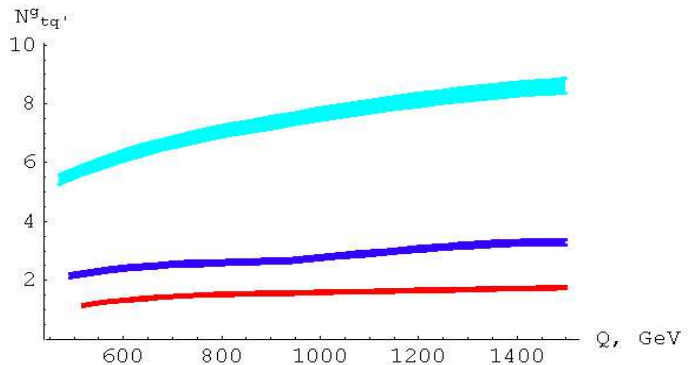}
}
\caption{\label{fig:Ngt2} T-channel single top production. Multiplicity $N^g_{tq'}(Q)$ 
versus $Q$ for different 
cuts of jet transverse 
momenta. Top-down: $P_t=10\;GeV\to P_t=30\;GeV\to P_t=50\;GeV$.}
\end{center} 
\end{figure}

% tab:Ngt2
\begin{table}[b!]
\caption{\label{tab:Ngt2} T-channel single top production. Multiplicity $N^g_{tq'}$ for different cuts of jet transverse momenta $P_t$ and
the energy of parton-parton collision.}
\centering
\begin{tabular}{|c|c|c|c|}
\hline
 {\cellcolor[gray]{0.9}$N^g_{tq'}(Q,P_t)$} & \multicolumn{3}{c|}{$Q$, GeV}  \\
 \hhline{|-|-|-|-|}
 $P_t$, GeV & 600 & 1000 & 1500 \\
 \hline
 10 & 6.23$\pm$0.18 & 7.65$\pm$0.22 & 8.62$\pm$0.24 \\
 \hline
 30 & 2.4$\pm$0.07 & 2.77$\pm$0.075 & 3.29$\pm$0.08 \\
 \hline
 50 & 1.32$\pm$0.038 & 1.59$\pm$0.04 & 1.76$\pm$0.044 \\
 \hline
\end{tabular}
\end{table}
 The process of t-channel single top production $pp\to t+X$ 
has higher rate (about 245~pb at 14~TeV) than the previous one, but we have to
make the same assumptions concerning fragmentation and color
reconnection processes as in $t\bar{t}$ production. Here
calculations for the parton level process~(\ref{prt2}) are 
presented. Numerical values 
for $N^g_{tq'}$ are given in the table~\ref{tab:Ngt2} and on 
the Fig.~\ref{fig:Ngt2}. 
% fig:Ngt2
The average charged multiplicity in different decay 
modes looks as follows
\begin{eqnarray}
 \label{mult:t2HH} N^h_{tq'\to hadrons}(Q)&\!=&\!n^H_t+n_q+N^g_{tq'}(Q),\\
\label{mult:t2HL} N^h_{tq'\to l\bar{\nu}_l+hadrons}(Q)&\!=&\!n^L_t+n_q+N^g_{tq'}(Q).
\end{eqnarray}
% tab:Ngt3
\begin{table}[t!]
\caption{\label{tab:Ngt3} Multiplicity $N^g_{tW}$ for different cuts of jet 
transverse momenta $P_t$ and
the energy of parton-parton collision.}
\centering
\begin{tabular}{|c|c|c|c|}
\hline
 {\cellcolor[gray]{0.9}$N^g_{tW}(Q,P_t)$} & \multicolumn{3}{c|}{$Q$, GeV}  \\
 \hhline{|-|-|-|-|}
 $P_t$, GeV & 600 & 1000 & 1500 \\
 \hline
 10 & 2.54$\pm$0.06 & 5.83$\pm$0.14 & 10.27$\pm$0.23 \\
 \hline
 30 & 0.85$\pm$0.017 & 2.27$\pm$0.04 & 4.49$\pm$0.076 \\
 \hline
 50 & 0.42$\pm$0.008 & 1.25$\pm$0.019 & 2.6$\pm$0.036 \\
 \hline
\end{tabular}
\end{table}
% fig:Ngt3
\begin{figure}[t!]
\begin{center} 
\resizebox{0.5\textwidth}{5cm}{%
  \includegraphics{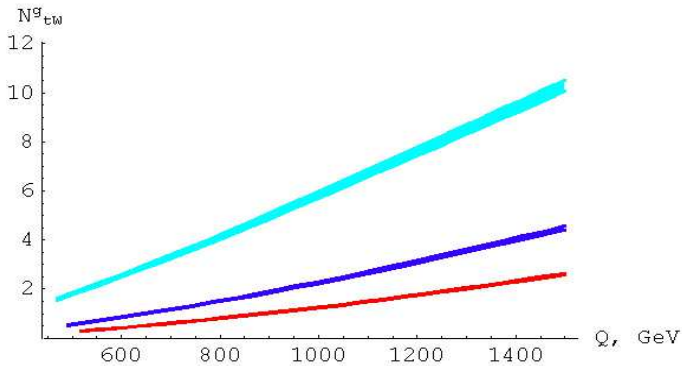}
}
\caption{\label{fig:Ngt3} Multiplicity $N^g_{tW}(Q)$ versus $Q$ for different cuts of 
jet transverse 
momenta. Top-down: $P_t=10\;GeV\to P_t=30\;GeV\to P_t=50\;GeV$. }
\end{center} 
\end{figure}
As you can see on the Fig.~\ref{fig:Ngt2}, the value of $N^g_{tq'}$ is rather
small in the wide
kinematical region, and energy dependence is not strong. \emph{\bf It is important for the 
estimation of the multiplicity from beam remnants plus color reconnection effects},
since values $n_t^H$, $n_t^L$, $n_q$ are fixed by previous measurements 
and $N^g_{tq'}(Q)\ll n_t$. From this point of view the t-channel single
top production looks the most interesting process for the multiplicity 
measurements.

$tW$ production has intermediate cross-section of the order 62~pb at 14~TeV which
lies between s- and t-channel single top production rates. Probably, specific
signature of this process would help in the measurements proposed 
in this work. Numerical 
values for $N^g_{tW}$ are given 
in the table~\ref{tab:Ngt3} and on 
the Fig.~\ref{fig:Ngt3}. The process~(\ref{prt3}) has 3 decay 
modes. The corresponding average charged multiplicities are
\begin{eqnarray}
 \label{mult:t3HH} N^h_{tW\to hadrons}(Q)&\!=&\!n^H_t+n_W+N^g_{tW}(Q),\\
 \label{mult:t3HL} N^h_{tW\to (W)l\bar{\nu}_l+hadrons}(Q)&\!=&\!n^H_t+N^g_{tW}(Q),\\
 \label{mult:t3LL} N^h_{tW\to l^+l^-\nu_l\bar{\nu}_l+hadrons}(Q)&\!=&\!n^L_t+N^g_{tW}(Q).
\end{eqnarray}
The energy dependence is also visible and can be used to test QCD calculations.

\section{Discussions and conclusions}
\label{conclusion}
\begin{figure}[b!]
\begin{center} 
\resizebox{0.5\textwidth}{7cm}{%
  \includegraphics{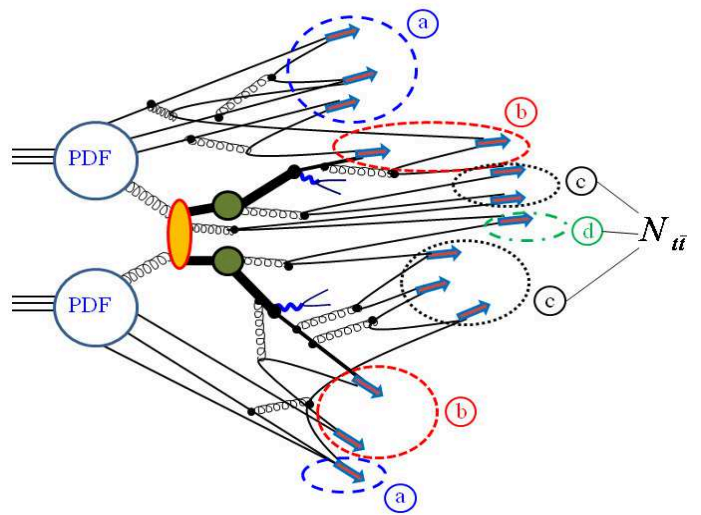}
}
\caption{\label{fig:fragm} Complicated fragmentation pattern of the inclusive $t\bar{t}$ production
in pp collisions. a) beam remnants; b) hadrons arising from the color interaction of beam 
remnants with
the final state radiation from quarks (suppressed for high-$P_t$ of final quarks or gluons); c) hadrons 
from top fragmentation; d) result of $t^*\bar{t}^*$ interaction.}
\end{center} 
\end{figure}
In this article we consider four processes with top production at the LHC. Average
charged hadronic multiplicities were calculated in perturbative QCD. Hadronic 
multiplicity in a gluon is fixed by low energy data.

There are several important tasks that could 
be solved by multiplicity measurements: 
to test QCD calculations \emph{\bf independently on fragmentation 
models}, to check independent fragmentation of heavy quarks, to 
check parton-parton C.M. energy dependence of hadron  
multiplicities, to estimate multiplicity from beam remnants plus 
from color reconnection effects in t-channel single top for 
further use in other processes. We can calculate also the difference
$\Delta N_{Qq}\equiv N_Q-N_q$ to cancel 
effects of color reconnection and beam remnants. 

There are several assumptions in the present work:
\begin{itemize}
\item independent fragmentation of on-shell top
quarks in $t\bar{t}$ production;
\item color reconnection effects in the interaction of jets with
beam remnants (for nonsinglet production of $t\bar{t}$, $tq'$, $tW$) 
are suppressed for large lower cuts in jet transverse momenta. As you
can see on the Fig.~\ref{fig:fragm}, the fragmentation pattern in $t\bar{t}$
production is rather complicated. We have beam remnants with
low transverse momenta interacting with 
jet remnants with large transverse momenta. Amplitudes
for such processes are suppressed for high $P_t$ since they are
propotional to the inverse power of the momentum
transfer squared $\hat{t}$ in the parton-parton interaction, and
\begin{eqnarray}
\hat{t}&\!=&\!(p^{beam}_0-p^{jet}_0)^2-(p^{beam}_3-p^{jet}_3)^2-\nonumber\\
&&(\vec{p}^{\;beam}_t-\vec{p}^{\;jet}_t)^2
\simeq -{p^{jet}_t}^2<-P_t^2,\nonumber
\end{eqnarray}
where $p^{beam}$ and $p^{jet}$ are momenta of beam and 
jet partons correspondingly, and from the kinematics
$$
p^{beam}_0\sim |p^{beam}_3|\gg p^{jet}_0\sim |p^{jet}_3|,\ p^{jet}_t\gg p^{beam}_t.
$$ 
We are  interested only in c and d types of fragmentation depicted
on the Fig.~\ref{fig:fragm}.
\item in the s-channel $t\bar{b}$ singlet production there is no
color reconnection with beam remnants.
\end{itemize}
All the above assumptions are based on low energy data and
theoretical estimations of pQCD and also can be\linebreak checked
at the LHC.

From the experimental point of view t-channel single top production~(\ref{prt2})
is the most convenient case, since the energy dependence of the
average charged hadronic multiplicity is weak. We can \emph{\bf estimate 
quantitatively effect of color reconnection of jets with beam remnants} to
check our assumption on its suppression. Then
we can use this estimation to improve our predictions for other channels
of top production. For this task it is also useful to extract multiplicities
in different decay modes of top quarks. 

The final experimental 
task is to extract number of tracks in jets which are produced in top quark
decays. To estimate experimental efficiencies and dependence on 
a fragmentation model
we can use any MC generator for top production. At 
the same time with the top-mass reconstruction procedure (in hadronic mode) we could extract number of tracks which are included into 
hadronic cluster from single top or top anti-top decays. At the 
moment we have
a good chance to make the new independent test of QCD 
by the use of
recent LHC data at $7$~TeV. Other experimental aspects of such
measurements will be discussed in futher works.

\section*{Appendix A}
\label{appA}
Let us consider the process 
\begin{equation}
parton_1(q_1)+parton_2(q_2)\to c(p_1)+d(p_2)+g(k)
\label{2to3proc}
\end{equation}
and 
put $q_i^2=0$, $k^2=K^2>0$ (we put also $m_b=0$ for 
processes~(\ref{prt3}),(\ref{prt4}), since corrections are
of the order $m_b^2/m_t^2\ll 1$). In the C.M. frame
of colliding partons we can write:
\begin{eqnarray}
&& q_1=\frac{Q}{2}\left( 1,0,0,1\right),\; q_2=\frac{Q}{2}\left( 1,0,0,-1\right),\nonumber\\
&& q=q_1+q_2,\; \Delta=(q_1-q_2)/2,\label{eq:kin1}\\
&& k=\left( \frac{qk}{Q},\frac{\sqrt{D}}{Q}\sin\theta_k,0,\frac{\sqrt{D}}{Q}\cos\theta_k\right),\nonumber\\
&& l=\left( \frac{Q^2+Z}{2Q},|{\vec{l}\;\!}|\sin\theta_l\cos\phi,|{\vec{l}\;\!}|\sin\theta_l\sin\phi,|{\vec{l}\;\!}|\cos\theta_l\right),\nonumber\\
&& p_1=l-k,\; p_2=q-l,\label{eq:kin2}
\end{eqnarray}
\begin{eqnarray}
&& D=(qk)^2-Q^2K^2,\; |{\vec{l}\;\!}|=\sqrt{\frac{\left( Q^2-Z\right)^2}{4Q^2}-m_2^2},\nonumber\\
&& \hat{s}=(q-k)^2=Q^2+K^2-2qk,\; Z=l^2-m_2^2,\label{eq:kin3}\\
&& \cos\theta_{kl}=\cos\theta_k\cos\theta_l+\sin\theta_k\sin\theta_l\cos\phi\;\;\;\;\;\mbox{\rm or}\nonumber\\
&& \label{momenta2to3}\cos\theta_l=\cos\theta_{kl}\cos\theta_k+\sin\theta_{kl}\sin\theta_k\cos\phi^*.
\end{eqnarray}

After change of variables phase space looks as follows:
\begin{eqnarray}
&& \int dK^2 \int d\Phi_{2\to3}=\int dK^2\int\!\!\!\!\int\frac{d^4k}{(2\pi)^4}\frac{d^4l}{(2\pi)^4}\times\nonumber\\
&& (2\pi)\delta(p_1^2-m_1^2)(2\pi)\delta(p_2^2-m_2^2)(2\pi)\delta(K^2-k_{\mu}k^{\mu})=\nonumber\\
&& \left[ \frac{\sqrt{D}}{2(2\pi)^2Q^2}\int\!\!\!\!\int\!\!\!\!\int dK^2\; d(qk)\;d\cos\theta_k\right]\times\nonumber\\
&& \left[ \int\!\!\!\!\int\!\!\!\!\int\frac{\delta(\mathcal{H}(Z))}{2(4\pi)^2\sqrt{D}}\; dZ\; d\phi\; d\cos\theta_l\right]=\nonumber\\
&& \frac{1}{2(4\pi)^3 Q^2}\int\limits_{K_-^2}^{K_+^2}dK^2\int\limits_{(qk)_-}^{(qk)_+}d(qk)\int\limits_{C_{k,-}}^{C_{k,+}}d\cos\theta_k\int\limits_{Z_-}^{Z_+}dZ\int\limits_0^{\pi}\frac{d\phi}{\pi},\nonumber\\
&& \!\!\!\mathcal{H}(Z)=\cos\theta_{kl}(Z)-\cos\theta_k\cos\theta_l-\sin\theta_k\sin\theta_l\cos\phi.\label{phspace2to3} 
\end{eqnarray}
Here we keep the integration in $K^2$ since the gluon is virtual.

Then we have to cut jet transverse momenta from below to
suppress color reconnection with beam remnants
$$
p_{i,\perp}\ge P_{t,i},\; k_{\perp}\ge K_t,
$$
or
$$
p_{i,3}\le\sqrt{\vec{p}_i^2-P_{t,i}^2},\; k_3\le\sqrt{\vec{k}^2-K_t^2}. 
$$
In this paper $K_t=P_{t,i}=P_t$. Finally we have conditions
\begin{eqnarray}
&&\label{Tcondition1} \left|\cos\theta_l\right|\left|\vec{l}\right|\le\sqrt{\vec{l}^2-P_{t,2}^2}, \\
&&\label{Tcondition2} \left| \left|\vec{l}\right|\cos\theta_l-\left|\vec{k}\right|\cos\theta_k \right|\le\sqrt{\left( \vec{l}-\vec{k}\right)^2-P_{t,1}^2}.
\end{eqnarray}

For limits in the above integrals without conditions~(\ref{Tcondition1}),(\ref{Tcondition2}) we can write
\begin{eqnarray}
&& K_-^2=Q_0^2,\; K_+^2=\left(Q-\sqrt{(m_1+m_2)^2+K_t^2}\right)^2-K_t^2,\nonumber\\
&& (qk)_-=Q\sqrt{K^2+K_t^2},\nonumber\\
&& (qk)_+=\frac{Q^2+K^2-(m_1+m_2)^2}{2},\nonumber\\
&& m_{i,\perp}=\sqrt{m_i^2+P_t^2},\nonumber\\
&& C_{k,\pm}=\pm \sqrt{\frac{D-Q^2K_t^2}{D}},\nonumber\\
&& Q\ge\sqrt{(m_1+m_2)^2+K_t^2}+\nonumber\\
&& \phantom{Q\ge}\;+\sqrt{Q_0^2+K_t^2}> m_1+m_2+Q_0,\nonumber\\
&& Q\le x_{i,max}\sqrt{s}.\label{klimits}
\end{eqnarray}
Taking into account the inequality ($\cos\theta_{kl}$ can be obtained from $\delta((k-l)^2-m_1^2)$)
\begin{equation}
\label{Zinequality}
\left|\cos\theta_{kl}\right|=\left| -\frac{A_l}{\frac{Q}{2}|{\vec{l}\;\!}|}\right|\le 1,
\end{equation}
where 
\begin{equation}
\label{coefAl}
A_l=\frac{Q^2}{4\sqrt{D}}\left[
K^2-qk+Z\left( 1-\frac{qk}{Q^2}\right)+m_2^2-m_1^2
\right]
\end{equation}
and $|\vec{l}|$(see~(\ref{momenta2to3})) depend on $Z$,
we can obtain limits:
\begin{eqnarray}
&&\!\!\!\! Z_{\pm}=\nonumber\\
&&\!\! qk+\frac{(Q^2-qk)(m_1^2-m_2^2)}{\hat{s}}\pm\sqrt{D}\sqrt{D_{12}(\hat{s},m_1,m_2)},\nonumber\\
&& \!\!\!\!D_{12}(s,m_1,m_2)=\nonumber\\
&& \phantom{D_{12}}\!\left( 1-\frac{(m_1+m_2)^2}{s}\right)\left(1-\frac{(m_1-m_2)^2}{s} \right).
\label{intlimits}
\end{eqnarray}
For multidimensional integration it is convenient to
introduce undimensional variables and make appropriate symmetrization of the
function under the integration:
\begin{eqnarray}
&& \phi=x_{\phi}\pi,\;x_{\phi}\in[0,1],\nonumber\\
&& Z=qk+\frac{(Q^2-qk)(m_1^2-m_2^2)}{\hat{s}}+\nonumber\\
&& \zeta\sqrt{D}\sqrt{D_{12}(\hat{s},m_1,m_2)},\;\zeta\in[-1,1],\nonumber\\ 
&& \cos\theta_k=\tau\sqrt{\frac{D-Q^2K_t^2}{D}},\;\tau\in[-1,1],\nonumber\\
&& qk=(qk)_-+\eta((qk)_+-(qk)_-),\;\eta\in[0,1],
\label{undimvars}
\end{eqnarray}
\begin{eqnarray}
&& \label{symint}\int d(qk)\;d\cos\theta_k\; dZ\; \frac{d\phi}{\pi}\; f(qk,\cos\theta_k,Z,\phi)=\nonumber\\
&& \int\limits_0^1 d\eta\;d\tau\;d\zeta\;dx_{\phi}\; \mathcal{D}\;\tilde{f}^{sym},
\end{eqnarray}
\begin{eqnarray}
\label{symfun}\tilde{f}^{sym}&\!=&\!\tilde{f}(\eta,\tau,\zeta,x_{\phi})+\tilde{f}(\eta,-\tau,\zeta,x_{\phi})+\nonumber\\
&& \tilde{f}(\eta,\tau,-\zeta,x_{\phi})+\tilde{f}(\eta,-\tau,-\zeta,x_{\phi}),\\
   \mathcal{D}&\!=&\!((qk)_+-(qk)_-)\times\nonumber\\
&& \sqrt{D-Q^2K_t^2}\sqrt{D_{12}(\hat{s},m_1,m_2)},\label{detundim}
\end{eqnarray}
where $\tilde{f}$ is equal to $f$ after the change of variables.

In this paper we consider the case, when the final gluon jet can not
be separated experimentally from one of final quark jets:
$$
\cos\theta_{gq}>R=0.9,\;\mbox{or}\; \theta_{gq}<0.45.
$$
The above inequality leads to the following conditions
\begin{eqnarray}
\label{Rcondition1} \cos\theta_{\vec{p}_2\vec{k}}&=&-\cos\theta_{kl}>R,\\
\label{Rcondition2}\mbox{or}\; \cos\theta_{\vec{p}_1\vec{k}}&=&\left.-\cos\theta_{kl}\right|_{\zeta\to -\zeta}>R, 
\end{eqnarray}
where $\cos\theta_{kl}$ is expressed in terms of variables $\zeta$, $\eta$, $x_{\phi}$, $\tau$.

Let us denote 
conditions~(\ref{Tcondition1}),(\ref{Tcondition2}),(\ref{Rcondition1}),(\ref{Rcondition2}) as a product of corresponding $\theta$-functions
\begin{eqnarray}
&& f_{T,R}=\theta\left( \sqrt{\vec{l}^2-P_{t,2}^2}-
\left|\cos\theta_l\right|\left|\vec{l}\right|
\right)\times\nonumber\\
&& \theta\left(
\sqrt{\left( \vec{l}-\vec{k}\right)^2-P_{t,1}^2}-
\left| \left|\vec{l}\right|\cos\theta_l-\left|\vec{k}\right|\cos\theta_k \right|
\right)\times\nonumber\\
&&\label{condfunTR} \left[\theta\left(
-\cos\theta_{kl}-R
\right)\;\mbox{or}\; \theta\left(
-\left.\cos\theta_{kl}\right|_{\zeta\to -\zeta}-R
\right)\right],
\end{eqnarray}
where
$$
\theta(x)=\left\{ {1,\; x\ge 0}\atop {0,\; x<0} \right.\;.
$$
Now we can rewrite the second term in the r.h.s. of Ref.(\ref{hadron_mult2}) as follows
\begin{eqnarray}
&& \int \frac{d^4k}{(2\pi)^4} \,
\Pi(Q^2,k^2,kq_1,kq_2) f_{T,R}\times\nonumber\\
&& \label{hadron_mult3}
\,\,\,\,\;\left.\frac{d}{dp^2}N_g(p^2)\right|_{p^2=k^2}=\frac{I_1}{I_2},\\
&& \label{hadron_mult3a}I_1=\frac{1}{2(4\pi)^2 Q^2}\int\limits_{K_-^2}^{K_+^2}dK^2\; \alpha_s(K^2)\left.\frac{d}{dp^2}N_g(p^2)\right|_{p^2=K^2}\times \nonumber\\
&& \,\,\,\,\;\int\limits_{(qk)_-}^{(qk)_+}d(qk)\int\limits_{C_{k,-}}^{C_{k,+}}d\cos\theta_k\int\limits_{Z_-}^{Z_+}dZ\int\limits_0^{\pi}\frac{d\phi}{\pi}\; f_{T,R}\;\tilde{\Pi}_{2\to 3},\\
&& \label{hadron_mult3b} I_2=\frac{\sqrt{D_{12}(Q^2,m_1,m_2)}}{16\pi}\int\limits_{-C'}^{+C'}d\cos\theta_{q_1p_1}\tilde{\Pi}_{2\to 2},\\
&& C'^2=\frac{D_{12}(Q^2,m_{1,\perp},m_{2,\perp})}{D_{12}(Q^2,m_1,m_2)},
\end{eqnarray}
where $\tilde{\Pi}_{2\to 3}$($\tilde{\Pi}_{2\to 2}$) is the amplitude squared of 
the corresponding process~(\ref{prtt})-(\ref{prt3}) 
with (without) gluon radiation, which is calculated in Appendices~B,C. For simplicity
we put all the coupling constants to unity in these quantities. Here
$g_s=\sqrt{4\pi\alpha_s}$ is 
the QCD coupling constant. All tensors are 
contracted as in~(\ref{convolution}),(\ref{PIfun}).  

Different kinematical invariants of the process~(\ref{2to3proc}) can be expressed 
in terms of $Q^2$, $K^2$, $qk$, $\cos\theta_k$, $Z$, $\phi$ (and then $Q^2$, $K^2$, $\eta$, $\tau$, $\zeta$, $x_{\phi}$):
\begin{eqnarray}
&& q_1q_2=\frac{Q^2}{2},\; p_1p_2=\frac{\hat{s}-m_1^2-m_2^2}{2},\;\nonumber\\
&& p_1q_1=\frac{Q^2+Z-2qk}{4}+\Delta l-\Delta k,\nonumber\\
&& p_1q_2=\frac{Q^2+Z-2qk}{4}-\Delta l+\Delta k,\nonumber\\
&& p_2q_1=\frac{Q^2-Z}{4}-\Delta l,\; p_2q_2=\frac{Q^2-Z}{4}+\Delta l,\nonumber
\end{eqnarray}
\begin{eqnarray}
&& p_1k=\frac{Z-K^2-m_1^2+m_2^2}{2},\nonumber\\
&& p_2k=\frac{2qk-Z-K^2+m_1^2-m_2^2}{2},\nonumber\\
&& q_1k=\frac{qk}{2}+\Delta k,\; q_2k=\frac{qk}{2}-\Delta k,\nonumber\\
&& \Delta k=-\frac{\sqrt{D}}{2}\cos\theta_k,\nonumber\\
&& \Delta l=A_l\cos\theta_k-B_l\sin\theta_k\cos\phi,
\end{eqnarray}
\begin{eqnarray}
 && \!\!\!\!A_l=\nonumber\\
&& \!\!\!\!\zeta\frac{Q^2-qk}{4}\sqrt{D_{12}(\hat{s},m_{1,\perp},m_{2,\perp})}-\frac{\sqrt{D}}{4}\left( 1-\frac{m_1^2-m_2^2}{\hat{s}}\right),\nonumber\\
 && \!\!\!\!B_l=\nonumber\\
 &&\!\!\!\!\!\sqrt{\frac{Q^2}{4}|{\vec{l}\;\!}|^2-A_l^2}=\frac{Q\sqrt{\hat{s}}}{4}\sqrt{D_{12}(\hat{s},m_{1,\perp},m_{2,\perp})}\sqrt{1-\zeta^2}.
\label{inariants2to3}
\end{eqnarray}

\section*{Appendix B}
\label{AppB}
 Here we consider amplitudes for the top-antitop production~(\ref{prtt}).
% fig:tt 
\begin{figure}[tb!]
\begin{center} 
\resizebox{0.4\textwidth}{3cm}{%
  \includegraphics{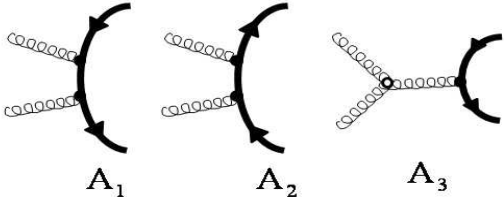}
}
\caption{\label{fig:tt} Amplitudes of the process $gg\to Q\bar{Q}$.}
\end{center} 
\end{figure}
 For the amplitude of the process without additional gluon radiation we 
 have three diagrams of Fig.~\ref{fig:tt}, and $I_2$ can be calculated as follows
 \begin{eqnarray}
  \!\!\!I_2&\!=&\!\frac{8}{3\pi} \biggl\{
 8(1+4x_m^2+x_m^4)\ln\frac{1+\sqrt{1-4x_{m\perp}^2}}{2x_{m\perp}}+\nonumber\\
 \!\!\! && \!\!\!\!\sqrt{1-4x_{m\perp}^2}\left( 
 3x_{m\perp}^2-18x_m^2-7-16\frac{x_m^4}{x_{m\perp}^2})
 \right)
\biggr\},\label{TT2to2ttbar}
\end{eqnarray}
\begin{equation}
  \label{TT2to2ttbar:a}x_m=m_t/Q,\; x_{m\perp}=m_{t,\perp}/Q.
 \end{equation}

%fig:ttg
\begin{figure}[tb!]
\begin{center} 
\resizebox{0.5\textwidth}{5cm}{%
  \includegraphics{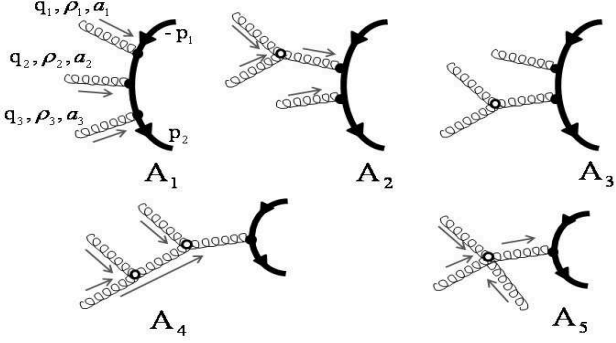}
}
\caption{\label{fig:ttg} $gggQ\bar{Q}$ amplitudes.}
\end{center} 
\end{figure}
For the amplitude of the process with additional gluon we 
have five kinds of diagrams (see Fig.~\ref{fig:ttg}):
 \begin{eqnarray}
\!\!\!&& A_i^{123}=\bar{u}(p_2,m_t)\hat{A}_i^{123}v(p_1,m_t),\\ 
\!\!\!&& \label{ttgA1} \hat{A}_{1}^{123}=[321]\frac{\gamma_{\rho_3}\left( \hat{p}_2-\hat{q}_3+m_t\right)\gamma_{\rho_2}\left(\hat{q}_1-\hat{p}_1+m_t \right)\gamma_{\rho_1}}{(q_3^2-2p_2q_3)(q_1^2-2p_1q_1)},\nonumber\\
\!\!\!&& \hat{A}_{2}^{123}=\left( [321]-[312]\right)
\frac{\gamma_{\rho_3}\left( \hat{p}_2-\hat{q}_3+m_t\right)\gamma_{\lambda}}{(q_3^2-2p_2q_3)}
\frac{\tilde{d}^{\lambda\beta}(q_1+q_2)}{(q_1+q_2)^2}\times\nonumber\\
\!\!\!\!\!\!\!&& \label{ttgA2}\left\{ (q_1-q_2)_{\beta}g_{\rho_1\rho_2}-(2q_1+q_2)_{\rho_2}g_{\beta\rho_1}+(2q_2+q_1)_{\rho_1}g_{\beta\rho_2}\right\},\nonumber
\end{eqnarray}
\begin{eqnarray}
\!\!\!\!\!&& \hat{A}_{3}^{123}=\left( [213]-[123]\right)
\frac{\gamma_{\lambda}\left( \hat{q}_3-\hat{p}_1+m_t\right)\gamma_{\rho_3}}{(q_3^2-2p_1q_3)}
\frac{\tilde{d}^{\lambda\beta}(q_1+q_2)}{(q_1+q_2)^2}\times\nonumber\\
\!\!\!\!\!&& \label{ttgA3}\left\{ (q_1-q_2)_{\beta}g_{\rho_1\rho_2}-(2q_1+q_2)_{\rho_2}g_{\beta\rho_1}+(2q_2+q_1)_{\rho_1}g_{\beta\rho_2}\right\},\nonumber\\
\!\!\!\!\!&& \hat{A}_{4}^{123}=\left( [312]+[213]-[321]-[123]\right)
\gamma_{\lambda}\frac{\tilde{d}^{\lambda\lambda'}(p_1+p_2)}{(p_1+p_2)^2}
\times\nonumber\\
\!\!\!\!\!&& \left\{ (2q_3-(p_1+p_2))_{\lambda'}g_{\rho_3\beta'} +(2(p_1+p_2)-q_3)_{\rho_3}g_{\lambda'\beta'}-\right.\nonumber\\
\!\!\!\!\!&& \left.(q_3+(p_1+p_2))_{\beta'}g_{\rho_3\lambda'}\right\}
\frac{\tilde{d}^{\beta'\beta}(q_1+q_2)}{(q_1+q_2)^2} \times\nonumber\\
\!\!\!\!\!&& \label{ttgA4} \left\{ (q_1-q_2)_{\beta}g_{\rho_1\rho_2}-(2q_1+q_2)_{\rho_2}g_{\beta\rho_1}+(2q_2+q_1)_{\rho_1}g_{\beta\rho_2}\right\}\nonumber\\
\!\!\!\!\!&& \hat{A}_{5}^{123}=\gamma_{\lambda}\frac{\tilde{d}^{\lambda\beta}(p_1+p_2)}{(p_1+p_2)^2}\times\nonumber\\
\!\!\!\!\!&&\left\{ 
\left( [312]+[213]\right)\left( g_{\beta\rho_2}g_{\rho_1\rho_3}+g_{\beta\rho_3}g_{\rho_1\rho_2}-2g_{\beta\rho_1}g_{\rho_2\rho_3}\right)+
\right.\nonumber\\
\!\!\!\!\!&& \left.  
\left( [321]+[123]\right)\left( g_{\beta\rho_1}g_{\rho_2\rho_3}+g_{\beta\rho_3}g_{\rho_1\rho_2}-2g_{\beta\rho_2}g_{\rho_1\rho_3}\right)+
\right.\nonumber\\
\!\!\!\!\!&& \label{ttgA5}\left. 
\left( [132]+[231]\right)\left( g_{\beta\rho_1}g_{\rho_2\rho_3}+g_{\beta\rho_2}g_{\rho_1\rho_3}-2g_{\beta\rho_3}g_{\rho_1\rho_2}\right)
\right\},
\end{eqnarray}
where we consider all gluons as initial particles. Then we can calculate
\begin{eqnarray}
&&\!\!\!\!\!\!\!\!\!\! T_{2\to3\; \rho_1\rho_2\rho_3}^{\;\;\;\;\;\;\; a_1a_2a_3}(q_1,q_2,q_3)=\nonumber\\
 && \;\biggl\{ A_1^{123}+A_1^{132}+A_1^{213}+A_1^{231}+A_1^{312}+A_1^{321}+\nonumber\\
&& \;\;\;\label{T2to3ttbarg}
\sum_{i=2}^{4}\left( A_i^{123}+A_i^{132}+A_i^{231} \right)+A_5^{123}
\biggr\},\\
&&\!\!\!\!\!\!\!\!\!\!\Pi^{\;\;\;\;\;\;\;\{ a_ib_i\};ab}_{2\to3\; \{ \rho_i\sigma_i\};\mu\nu}=\nonumber\\
&& \,\label{TT2to3ttbarg}T_{2\to3\; \rho_1\rho_2\mu}^{\;\;\;\;\;\;\; a_1a_2a}(q_1,q_2,-k)T_{2\to3\; \sigma_1\sigma_2\nu}^{*\;\;\;\;\;\; b_1b_2b}(q_1,q_2,-k).
\end{eqnarray}
Here $q_3=-k$ since \emph{we have one gluon in the final state} with momentum $k$,
 $A_i^{123}\equiv A_{i\;\rho_1\rho_2\rho_3}^{\;\;a_1a_2a_3}(q_1,q_2,q_3,p_{1,2})$, \linebreak $[ijk]=t^{a_i}t^{a_j}t^{a_k}$, $t^a$ are SU(3) matrices, $u$ and $v$ are
Dirac spinors, $\gamma_{\rho}$ are Dirac matrices, $\hat{p}\equiv p^{\mu}\gamma_{\mu}$. 

If we apply contractions~(\ref{convolution}),(\ref{PIfun})
to~(\ref{TT2to3ttbarg}) and take into account the 
theorem~(\ref{GIQCD}) (it was checked  by direct calculations for~(\ref{T2to3ttbarg}) ) then we obtain $g_s^6\tilde{\Pi}_{2\to3}$
for the process~(\ref{prtt}). Since the final expression for
$\tilde{\Pi}_{2\to3}$ is very complicated, we evaluate it 
numerically. To get the final result for 
the $t\bar{t}$ multiplicity induced by gluon 
radiation ($N^g_{t\bar{t}}$ on the Fig.~\ref{fig:pr}a)
we have to substitute $I_2$ and $\tilde{\Pi}_{2\to 3}$ for
this process to~(\ref{hadron_mult3})-(\ref{hadron_mult3a}).

\section*{Appendix C}
\label{AppC}
For simplicity here we set all
coupling constants to unity. In this
section we consider calculations for 
processes~(\ref{prt1})-(\ref{prt3}).

%fig:Fdiags
\begin{figure}[htb!]
\begin{center} 
\resizebox{0.5\textwidth}{3cm}{%
  \includegraphics{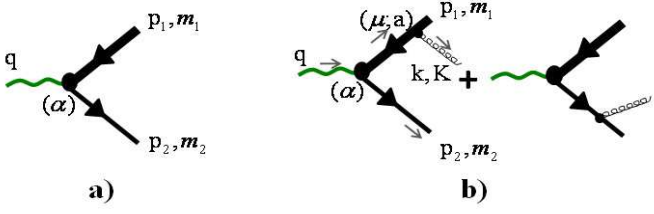}
}
\caption{\label{fig:Fdiags}Amplitudes for calculation of functions $\mathcal{F}^{(0)}$ (a) and $\mathcal{F}^{(1)}$ (b).}
\end{center} 
\end{figure}
Let us introduce some functions for futher
calculations. One of the functions is
the $WQ\bar{q}$ vertex squared (see Fig.~\ref{fig:Fdiags}a)
\begin{equation}
\mathcal{A}^{(0)}_{\alpha}=\bar{u}(p_2,m_2)\gamma_{\alpha}(1-\gamma_5)
v(p_1,m_1)\label{A0al},
\end{equation}
\begin{eqnarray}
&& \mathcal{F}^{(0)}_{\alpha\alpha'}(p_1,m_1,p_2,m_2)=
C_A\mathcal{A}^{(0)}_{\alpha}\mathcal{A}^{(0)\;*}_{\alpha'}=\nonumber\\
&&8C_A\biggl(  g_{\alpha\alpha'}p_1p_2
-p_{1\;\alpha}p_{2\;\alpha'}-p_{2\;\alpha}p_{1\;\alpha'}-\nonumber\\
&&\label{F0alal}\phantom{8C_A\biggr(} i\epsilon_{\alpha\alpha'\rho\sigma}p_{1\;\rho}p_{2\;\sigma}
\biggl).
\end{eqnarray}
The next one is the squared amplitude of the process $W\to Q\bar{q}g$ which is
shown on the Fig.~\ref{fig:Fdiags}b.
\begin{eqnarray}
&& \label{A1almu}\mathcal{A}^{(1)}_{\alpha\mu}=\bar{u}(p_2,m_2)\left\{
\frac{\gamma_{\alpha}(1-\gamma_5)\left(-\hat{p}_1-\hat{k}+m_1\right)\gamma_{\mu}}{K^2+2p_1k}+\right.\nonumber\\
&& \hspace*{1cm}\left.\frac{\gamma_{\mu}\left(\hat{p}_2+\hat{k}+m_2\right)\gamma_{\alpha}(1-\gamma_5)}{K^2+2p_2k}
\right\}v(p_1,m_1),\\
&& \label{F1alalmumu}
\mathcal{F}^{(1)}_{\alpha\alpha'\;\mu\mu'}(p_1,m_1,p_2,m_2,k,K)=
C_AC_F\mathcal{A}^{(1)}_{\alpha\mu}\mathcal{A}^{(1)\;*}_{\alpha'\mu'},\\
 && \tilde{\mathcal{F}}^{(1)}_{\alpha\alpha'}(p_1,m,p_2,0,k,K)=\nonumber\\
&&\mathcal{F}^{(1)}_{\alpha\alpha'\;\mu\mu'}(p_1,m,p_2,0,k,K)\left( -g^{\mu\mu'}+\frac{k^{\mu}k^{\mu'}}{K^2}\right)=\nonumber\\
&& \label{F1alal}
\frac{128}{\left( K^2+2p_1k\right)^2\left( K^2+2p_2k\right)^2}\sum\limits_{i=1}^{9}\tilde{\mathcal{F}}^{(1);\; i}_{\alpha\alpha'},
\end{eqnarray}
where
\begin{eqnarray}
&& \hspace*{-0.8cm}\tilde{\mathcal{F}}^{(1);\; 1}_{\alpha\alpha'}=g_{\alpha\alpha'}\times\nonumber\\
&&\hspace*{-0.7cm} \biggl\{ 
Q^4\left[
Z(2qk-Z)+2m^2(Z-qk)-m^4
\right]+
\nonumber\\
&&\hspace*{-0.5cm} 
Q^2\left[
-2K^2\left( 
m^2+qk-Z
\right)^2-2qk(2qk-Z)Z-
\right.
\nonumber\\
&&\hspace*{0.3cm}\left.
m^2(4qk-Z)Z+2m^4(qk-Z)+m^6
\phantom{2K^2\left( 
m^2+qk-Z
\right)^2}\hspace*{-3.1cm}\right]+
\nonumber\\
&&\hspace*{-0.5cm} 
K^2m^2\left[
2qk^2-2Z\;qk+Z^2+
2m^2(qk-Z)+m^4
\right]+
\nonumber\\
&&\hspace*{-0.5cm} 
Z(2qk-Z)(2qk^2-2Z\;qk+Z^2)+\nonumber\\
&&\hspace*{-0.6cm} 2m^2(-2qk^3+8qk^2 Z-7Z^2qk+2Z^3)+
\nonumber\\
&&\hspace*{-0.5cm}
2m^4(-3qk^2+7Z\;qk-3Z^2)-4m^6(qk-Z)-m^8
\biggr\},
\end{eqnarray}
\begin{eqnarray}
&&\hspace*{-0.8cm} \tilde{\mathcal{F}}^{(1);\; 2}_{\alpha\alpha'}=p_{1\;\alpha}p_{1\;\alpha'}\times\nonumber\\
&& \hspace*{-0.7cm}\biggl\{
K^2\left[
-Z(2qk-Z)+2m^2(qk-Z)+m^4
\right]+\nonumber\\
&& \hspace*{-0.5cm}
Z(2qk-Z)^2-m^2(4qk^2-8Z\;qk+3Z^2)-\nonumber\\
&& \hspace*{-0.5cm}
m^4(4qk-3Z)-m^6
\biggr\},\\
&& \hspace*{-0.8cm}\tilde{\mathcal{F}}^{(1);\; 3}_{\alpha\alpha'}=p_{2\;\alpha}p_{2\;\alpha'}\times\nonumber\\
&& \hspace*{-0.7cm}\biggl\{
K^2\left[
-Z(2qk-Z)+2m^2(qk-Z)+m^4
\right]+\nonumber\\
&& \hspace*{-0.5cm}Z^2(2qk-Z)-m^2Z(4qk-3Z)+\nonumber\\
&& \hspace*{-0.5cm}m^4(2qk-3Z)+m^6
\biggr\},\\
&& \hspace*{-0.8cm}\tilde{\mathcal{F}}^{(1);\; 4}_{\alpha\alpha'}=\left(p_{1\;\alpha}p_{2\;\alpha'}+
p_{2\;\alpha}p_{1\;\alpha'}\right)\times\nonumber\\
&& \hspace*{-0.7cm}\biggl\{
Q^2\left[
-Z(2qk-Z)+2m^2(qk-Z)+m^4
\right]+
\nonumber\\
&& \hspace*{-0.5cm}K^2\left[
-Z(2qk-Z)+2m^2(qk-Z)+m^4+2qk^2
\right]+\nonumber\\
&& \hspace*{-0.5cm}qk(2qk-Z)Z+2m^2qk^2+m^4 qk\biggr\},\\
&& \hspace*{-0.8cm}\tilde{\mathcal{F}}^{(1);\; 5}_{\alpha\alpha'}=\left(p_{1\;\alpha}k_{\alpha'}+
k_{\alpha}p_{1\;\alpha'}\right)\times\nonumber\\
&& \hspace*{-0.7cm}\biggl\{
\frac{Q^2}{2}\left[
-Z(2qk-Z)+2m^2(qk-Z)+m^4
\right]+
\nonumber\\
&& \hspace*{-0.5cm}K^2\left[
-Z(qk-Z)+m^2(qk-2Z)+m^4
\right]+\nonumber\\
&& \hspace*{-0.5cm}
\frac{1}{2}Z(2qk-Z)^2-m^2(qk-2Z)(2qk-Z)-\nonumber\\
&& \hspace*{-0.5cm}
\frac{m^4}{2}(6qk-5Z)-m^6
\biggr\},\\
&& \hspace*{-0.8cm}\tilde{\mathcal{F}}^{(1);\; 6}_{\alpha\alpha'}=\left(p_{2\;\alpha}k_{\alpha'}+
k_{\alpha}p_{2\;\alpha'}\right)\times\nonumber\\
&& \hspace*{-0.7cm}\biggl\{
\frac{Q^2}{2}\left[
-Z(2qk-Z)+2m^2(qk-Z)+m^4
\right]+
\nonumber\\
&& \hspace*{-0.5cm}K^2\left[
2qk^2-3Z\;qk+Z^2+m^2(3qk-2Z)+m^4
\right]+\nonumber\\
&& \hspace*{-0.5cm}
\frac{1}{2}Z^2(2qk-Z)+m^2(4qk^2-5Z\;qk+2Z^2)-\nonumber\\
&& \hspace*{-0.5cm}
\frac{m^4}{2}(8qk-5Z)+m^6
\biggr\},\\
&& \hspace*{-0.8cm}\tilde{\mathcal{F}}^{(1);\; 7}_{\alpha\alpha'}=i\epsilon_{\alpha\alpha'\sigma\rho}p_1^{\sigma}p_2^{\rho}\times\nonumber\\
&& \hspace*{-0.7cm}\biggl\{
Q^2\left[ -Z(2qk-Z)+2m^2(qk-Z)+m^4\right]+\nonumber\\
&& \hspace*{-0.5cm}
K^2\left[ 2qk^2-3Z\;qk+Z^2+m^2(3qk-2Z)+m^4\right]+\nonumber\\
&& \hspace*{-0.5cm}
Z(2qk-Z)(qk-Z)+m^2(2qk^2+4Z\;qk-3Z^2)+\nonumber\\
&& \hspace*{-0.5cm}
m^4(3Z-qk)-m^6
\biggr\},\\
&& \hspace*{-0.8cm}\tilde{\mathcal{F}}^{(1);\; 8}_{\alpha\alpha'}=i\left( \epsilon_{\alpha\sigma\rho\lambda}\mathcal{P}_{\alpha'}-
\epsilon_{\alpha'\sigma\rho\lambda}\mathcal{P}_{\alpha}
\right) k^{\sigma}p_1^{\rho}p_2^{\lambda},\\
&& \hspace*{-0.8cm}\mathcal{P}_{\alpha}=\left( k-p_1+p_2\right)_{\alpha}\left( Z^2-2Z+m^4\right)+\nonumber\\
&& \hspace*{-0.8cm}\phantom{\mathcal{P}_{\alpha}=(}2qk(Z-m^2)p_{1\;\alpha}
\end{eqnarray}

%fig:t3
\begin{figure}[ht!]
\begin{center} 
\resizebox{0.5\textwidth}{4.5cm}{%
  \includegraphics{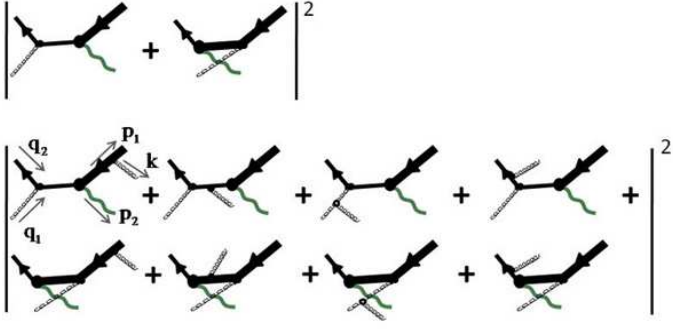}
}
\caption{\label{fig:t3} Diagrams for calculation of the process~(\ref{prt3}) and the tensor $\mathcal{F}^{(2)}$.}
\end{center} 
\end{figure}

\begin{eqnarray}
&& \hspace*{-0.8cm}\tilde{\mathcal{F}}^{(1);\; 9}_{\alpha\alpha'}=i\epsilon_{\alpha\alpha'\sigma\rho}k^{\sigma}p_2^{\rho}\times\nonumber\\
&& \hspace*{-0.7cm}\biggl\{ 
Q^2\left[ -Z(qk-Z)+m^2(qk-2Z)+m^4\right]+\nonumber\\
&& \hspace*{-0.5cm}
K^2\left[ 2qk^2-3Z\;qk+Z^2+m^2(3qk-2Z)+m^4\right]+\nonumber\\
&& \hspace*{-0.5cm}
m^2 \left[ 4qk^2-Z\;qk-Z^2+m^2(qk+2Z)-m^4\right]
\biggr\},
\end{eqnarray}
$$
Z=l^2=(p_1+k)^2=(q-p_2)^2,\; m=m_t.
$$

And the last one is the amplitude squared of the process 
which is depicted in the
lower Fig.~\ref{fig:t3}
\begin{eqnarray}
&&\hspace*{-1.2cm}\mathcal{F}^{(2)}_{\alpha\alpha'\;\mu\mu'\;\rho\rho'}(p_1,m_1,q_1,m'_1,q_2,m'_2,k,K)=\nonumber\\
&& \hspace*{0.4cm}\left( \sum\limits_{i=1}^8 \mathcal{A}^{(2)\;i}_{\alpha\mu\rho}\right)\left(\sum\limits_{j=1}^8\mathcal{A}^{(2)\;j}_{\alpha'\mu'\rho'}\right)^*,\; p_2^2=m_W^2.\label{F2alalmumuroro}  
\end{eqnarray}
Here $C_A=N=3$ and $C_F=(N^2-1)/(2N)=4/3$ are sructure constants of the group 
SU(3), color indices are contracted with $\delta_{aa'}\delta_{bb'}$ 
in~(\ref{F2alalmumuroro}). Expressions for Feinman diagrams looks as follows
\begin{eqnarray}
&& \mathcal{A}^{(2)\;i}_{\alpha\mu\rho}
=\bar{v}(q_2,m'_2) \hat{\mathcal{A}}^{(2)\;i}_{\alpha\mu\rho} v(p_1,m_1),\\
&& \label{A2almuro1}\hat{\mathcal{A}}^{(2)\;1}_{\alpha\mu\rho}
=
\frac{
\gamma_{\rho}\left( -\hat{q}+m'_2\right)\gamma_{\alpha}(1-\gamma_5)
\left( -\hat{p}_1-\hat{k}+m_1\right)\gamma_{\mu}
}{\left(Q^2-{m'_2}^2\right)\left( K^2-2p_1k\right)}t^at^b,\nonumber\\
&& \label{A2almuro2}
\hat{\mathcal{A}}^{(2)\;2}_{\alpha\mu\rho}=
\frac{
\gamma_{\rho}\left( -\hat{q}+m'_2\right)\gamma_{\mu}\left( \hat{k}-\hat{q}+m'_2\right)\gamma_{\alpha}(1-\gamma_5)
}{\left(Q^2-{m'_2}^2\right)\left((q-k)^2-{m'_2}^2\right)}t^at^b,\nonumber\\
&& \hat{\mathcal{A}}^{(2)\;3}_{\alpha\mu\rho}=
\frac{
\gamma_{\lambda}\left( \hat{k}-\hat{q}+m'_2\right)\gamma_{\alpha}(1-\gamma_5)
}{(q_1-k)^2\left((q-k)^2-{m'_2}^2\right)}\times\nonumber\\
&& \hspace*{1.5cm}\left( t^at^b-t^bt^a \right)\tilde{d}^{\lambda\beta}(q_1-k)\times\nonumber\\
&& \label{A2almuro3}
\hspace*{1.5cm}\left\{
\left( 2k-q_1\right)_{\rho}g_{\beta\mu}-\left(k+q_1\right)_{\beta}g_{\mu\rho}+
\left(2q_1-k\right)_{\mu}g_{\beta\rho}
\right\},\nonumber\\
&& \label{A2almuro4}
\hat{\mathcal{A}}^{(2)\;4}_{\alpha\mu\rho}=
\frac{
\gamma_{\mu}\left( \hat{k}-\hat{q}_2+m'_2\right)\gamma_{\rho}\left( \hat{k}-\hat{q}+m'_2\right)\gamma_{\alpha}(1-\gamma_5)
}{\left((q_2-k)^2-{m'_2}^2\right)\left((q-k)^2-{m'_2}^2\right)}t^bt^a,\nonumber
\end{eqnarray}
\begin{eqnarray}
&&\hspace*{-0.3cm} \label{A2almuro5}\hat{\mathcal{A}}^{(2)\;5}_{\alpha\mu\rho}
=
\frac{
\gamma_{\alpha}(1-\gamma_5)\left( \hat{p}_2-\hat{q}_2+m_1\right)\gamma_{\rho}
\left( -\hat{p}_1-\hat{k}+m_1\right)\gamma_{\mu}
}{\left((p_2-q_2)^2-m_1^2\right)\left( K^2-2p_1k\right)}t^at^b\!\!\!,\nonumber\\
&&\hspace*{-0.3cm} \label{A2almuro6}\hat{\mathcal{A}}^{(2)\;6}_{\alpha\mu\rho}
=
\frac{
\gamma_{\alpha}(1-\gamma_5)\left( \hat{p}_2-\hat{q}_2+m_1\right)\gamma_{\mu}
\left( \hat{q}_1-\hat{p}_1+m_1\right)\gamma_{\rho}
}{\left((p_2-q_2)^2-m_1^2\right)\left( (q_1-p_1)^2-m_1^2\right)}t^bt^a,\nonumber\\
&&\hspace*{-0.3cm} \hat{\mathcal{A}}^{(2)\;7}_{\alpha\mu\rho}=
\frac{
\gamma_{\alpha}(1-\gamma_5)\left( -\hat{q}_2+\hat{p}_2+m_1\right)\gamma_{\lambda}
}{(q_1-k)^2\left((p_2-q_2)^2-m_1^2\right)}\times\nonumber\\
&& \;\;\;\left( t^at^b-t^bt^a \right)\tilde{d}^{\lambda\beta}(q_1-k)\times\nonumber\\
&& \label{A2almuro7} 
\;\;\;\left\{
\left( 2k-q_1\right)_{\rho}g_{\beta\mu}-\left(k+q_1\right)_{\beta}g_{\mu\rho}+
\left(2q_1-k\right)_{\mu}g_{\beta\rho}
\right\},\nonumber\\
&&\hspace*{-0.3cm} \label{A2almuro8}\hat{\mathcal{A}}^{(2)\;8}_{\alpha\mu\rho}
=
\frac{
\gamma_{\mu}\left( \hat{k}-\hat{q}_2+m_1\right)\gamma_{\alpha}(1-\gamma_5)
\left( \hat{q}_1-\hat{p}_1+m_1\right)\gamma_{\rho}
}{\left((k-q_2)^2-{m'_2}^2\right)\left( (q_1-p_1)^2-m_1^2\right)}t^bt^a.\nonumber
\end{eqnarray}

Now we have all the ingredients to calculate amplitudes 
of processes~(\ref{prt1})-(\ref{prt3}). At first let 
us consider the s-channel single top production~(\ref{prt1}), which is shown
on the Fig.~\ref{fig:pr}b. We have to calculate $N^g_{t\bar{b}}$.

% fig:t1
\begin{figure}[b!]
\begin{center} 
\resizebox{0.5\textwidth}{4cm}{%
  \includegraphics{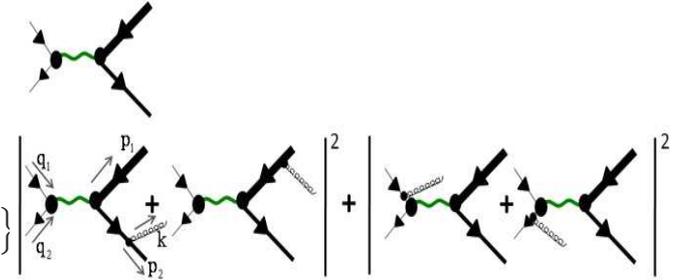}
}
\caption{\label{fig:t1} Diagrams for the calculation of the process~(\ref{prt1}).}
\end{center} 
\end{figure}
From upper and lower diagrams of the Fig.~\ref{fig:t1} we have
\begin{eqnarray}
&&  \hspace*{-0.8cm}\tilde{\Pi}_{2\to2}=\nonumber\\
&& \hspace*{-0.7cm}\label{TT2to2t1:a}d_W^{\alpha\beta}(q)d_W^{\alpha'\beta'}(q)\mathcal{F}^{(0)}_{\alpha\alpha'}(-q_1,0,-q_2,0)\mathcal{F}^{(0)}_{\beta\beta'}(p_1,m_t,p_2,0),\\
&& \hspace*{-0.3cm}I_2=\frac{6 \sqrt{D_{12}(Q^2,m_{t\;\perp},P_t)}}{\pi}\frac{1-x_m^2}{(1-x_w^2)^2}\times\nonumber\\
&& \hspace*{0.3cm}\label{TT2to2t1:b}\left(
3+C'^2+(3-C'^2)x_m^2\right),\\
&& \hspace*{-0.5cm}\label{TT2to2t1:c}C'^2=D_{12}(Q^2,m_{t\;\perp},P_t)/D_{12}(Q,m_t,0),\\
&& \hspace*{-0.4cm}x_w=m_W/Q,\nonumber 
\end{eqnarray}
and
\begin{eqnarray}
&& \tilde{\Pi}_{2\to 3}=\biggl\{
d_W^{\alpha\beta}(q)d_W^{\alpha'\beta'}(q)
\mathcal{F}^{(0)}_{\alpha\alpha'}(-q_1,0,-q_2,0)\times\nonumber\\
&& \phantom{\tilde{\Pi}_{2\to 3}=\biggl\{}\tilde{\mathcal{F}}^{(1)}_{\beta\beta'}(p_1,m_t,p_2,0,k,K)+\nonumber\\
&& \phantom{\tilde{\Pi}_{2\to 3}=\biggl\{}
d_W^{\alpha\beta}(q-k)d_W^{\alpha'\beta'}(q-k)
\mathcal{F}^{(0)}_{\beta\beta'}(p_1,m_t,p_2,0)\times\nonumber\\
&& \label{TT2to3t1}\phantom{\tilde{\Pi}_{2\to 3}=\biggl\{}\tilde{\mathcal{F}}^{(1)}_{\alpha\alpha'}(-q_1,0,-q_2,0,k,K)
\biggr\}
\end{eqnarray}
correspondingly, where 
$$
d^{\alpha\beta}_W(q)=\tilde{d}^{\alpha\beta}_W(q)/(q^2-m_W^2),\; \tilde{d}^{\alpha\beta}_W(q)=-g^{\alpha\beta}+\frac{q^{\alpha}q^{\beta}}{q^2},
$$
and for all the calculations we put $m_b=0$ since\linebreak $m_b/m_t\ll 1$.

\begin{figure}[t!]
\begin{center} 
\resizebox{0.5\textwidth}{4cm}{%
  \includegraphics{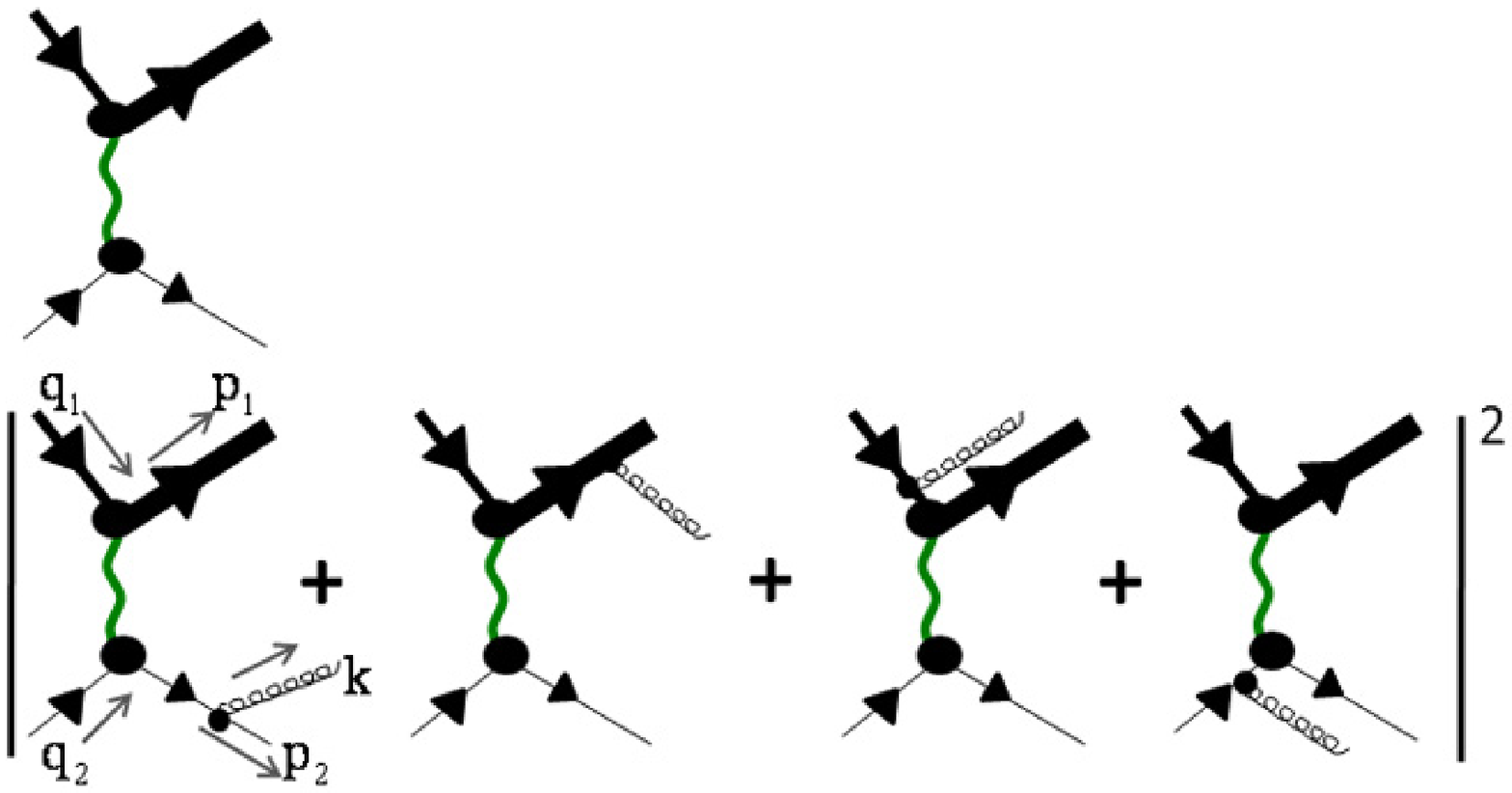}
}
\caption{\label{fig:t2} Diagrams for the calculation of the process~(\ref{prt2}).}
\end{center} 
\end{figure}

For the process~(\ref{prt2}) and calculation of $N^g_{tq'}$ we have the following functions (see diagrams on the Fig.~\ref{fig:t2})
\begin{eqnarray}
 \tilde{\Pi}_{2\to2}&\!=&\!d_W^{\alpha\beta}(q_1-p_1)d_W^{\alpha'\beta'}(q_1-p_1)\times\nonumber\\
&& \label{TT2to2t2:a}\mathcal{F}^{(0)}_{\alpha\alpha'}(p_2,0,-q_2,0)\mathcal{F}^{(0)}_{\beta\beta'}(p_1,m_t,-q_1,0),\\
I_2&\!=&\!\frac{288\sqrt{D_{12}(Q^2,m_{t\;\perp},P_t)}}{\pi}\times\nonumber\\
&&\label{TT2to2t2:b} \frac{1-x_m^2}{(1-x_m^2+2x_w^2)^2-C'^2(1-x_m^2)^2},
\end{eqnarray}
where $C'$ is the same as in the previous process.
\begin{eqnarray}
&& \hspace*{-0.8cm}\tilde{\Pi}_{2\to 3}=\nonumber\\
&& \hspace*{-0.8cm}\phantom{\tilde{\Pi}_{2\to 3}}\!\!\biggl\{
d_W^{\alpha\beta}(q_2-p_2)d_W^{\alpha'\beta'}(q_2-p_2)\times\nonumber\\
&& \hspace*{-0.5cm}\phantom{\tilde{\Pi}_{2\to 3}}\!\!\mathcal{F}^{(0)}_{\alpha\alpha'}(p_2,0,-q_2,0)\tilde{\mathcal{F}}^{(1)}_{\beta\beta'}(p_1,m_t,-q_1,0,k,K)+\nonumber\\
&& \hspace*{-0.5cm}\phantom{\tilde{\Pi}_{2\to 3}}\!\!
d_W^{\alpha\beta}(q_1-p_1)d_W^{\alpha'\beta'}(q_1-p_1)\times\nonumber\\
&& \hspace*{-0.5cm}\label{TT2to3t2}\phantom{\tilde{\Pi}_{2\to 3}}\!\!\mathcal{F}^{(0)}_{\beta\beta'}(p_1,m_t,-q_1,0)\tilde{\mathcal{F}}^{(1)}_{\alpha\alpha'}(p_2,0,-q_2,0,k,K)
\biggr\},
\end{eqnarray}
% fig:t2

Calculations for the process~(\ref{prt3}) and $N^g_{tW}$ looks as follows
(see diagrams on the Fig.~\ref{fig:t3})
\begin{eqnarray}
 \tilde{\Pi}_{2\to2}&\!=&\!
\left.\tilde{d}_W^{\alpha\alpha'}(p_2)\right|_{p_2^2=m_W^2}
\left.\tilde{d}^{\rho\rho'}(q_1,n)\right|_{n=\Delta}\times\nonumber\\
&& \label{TT2to2t3:a} \mathcal{F}^{(1)}_{\alpha\alpha'\;\rho\rho'}(p_1,m_t,-q_2,0,-q_1,0),
\end{eqnarray}
\begin{eqnarray}
&& I_2=\!\frac{1}{\pi}\Biggl\{
-2\sqrt{D_{12}(Q^2,m_{t\;\perp},m_{W\;\perp})}\times\nonumber\\
&& \biggl[u^2\left(1-C'^2\right)+2x_m^2u^2\left( 1+C'^2-u^2\left( 1-C'^2\right) \right)+
\nonumber\\
&& x_m^4u^2\left(1-u^2\right)^2\left( 1-C'^2\right)\biggr]^{-1}\times\nonumber\\
&& \biggl[
\left(3-2u^2\right)\left(1-C'^2\right)+
\nonumber\\
&& 
x_m^2\biggl( 3\left( C'^2+3\right)-u^2\left( C'^2+7\right)-2u^4 \left( 1-C'^2\right) \biggr)+
\nonumber\\
&& 
x_m^4\left(1-u^2\right) \biggl( 25+3C'^2+\nonumber\\
&& u^2\left( 23C'^2+33\right)-10u^4\left( 1-C'^2 \right) \biggr)+\nonumber\\
&&  3x_m^6\left( 1+2u^2\right)\left( 1-u^2\right)^3\left( 1-C'^2\right)\biggr]+
\nonumber\\
&&
\frac{4(1+2u^2)(1+2x_m^2(1-u^2)+2x_m^4(1-u^2)^2)}{u^2}\times\nonumber\\
&& \label{TT2to2t3:b}\ln\frac{1+x_m^2(1-u^2)+\sqrt{D_{12}(Q^2,m_{t\;\perp},m_{W\;\perp})}}{1+x_m^2(1-u^2)-\sqrt{D_{12}(Q^2,m_{t\;\perp},m_{W\;\perp})}}
\Biggr\},
\end{eqnarray}
where
\begin{eqnarray}
 C'^2&\!=&\!D_{12}(Q^2,m_{t\;\perp},m_{W\;\perp})/D_{12}(Q^2,m_t,m_W),\nonumber\\
 u&\!=&\!m_W/m_t,
\end{eqnarray}
\begin{eqnarray}
 \tilde{\Pi}_{2\to 3}&\!=&\!\left( -g^{\mu\mu'}+\frac{k^{\mu}k^{\mu'}}{K^2}\right)
\times\nonumber\\
&& \left.\tilde{d}_W^{\alpha\alpha'}(p_2)\right|_{p_2^2=m_W^2}\left.\tilde{d}^{\rho\rho'}(q_1,n)\right|_{n=\Delta}\times\nonumber\\
&&\label{TT2to3t3}\mathcal{F}^{(2)}_{\alpha\alpha'\;\mu\mu'\;\rho\rho'}(p_1,m_t,q_1,0,q_2,0,k,K).
\end{eqnarray}

\section*{Acknowledgements}

Author thanks V.A.~Petrov, A.V.~Kisselev, R.~Chierici, J.~Andrea and S.~Wimpenny
for fruitful discussions and useful comments.

%
% BibTeX users please use
% \bibliographystyle{}
% \bibliography{}

\end{document}